\documentclass[floatfix, noshowpacs, twocolumn, amsmath, amssymb, aps, prl, superscriptaddress]{revtex4}

\usepackage{bm}
\usepackage{amsmath}
\usepackage{amssymb}
\usepackage{graphicx}
\usepackage{xcolor}
\usepackage{bbold}
\usepackage{url}
\usepackage[caption=false]{subfig}

\usepackage{braket}
\usepackage{color}
\usepackage{times}

\begin{document}

\title{Experimental Scattershot Boson Sampling}

\author{Marco Bentivegna}
\affiliation{Dipartimento di Fisica, Sapienza Universit\`{a} di Roma,
Piazzale Aldo Moro 5, I-00185 Roma, Italy}

\author{Nicol\`o Spagnolo}
\affiliation{Dipartimento di Fisica, Sapienza Universit\`{a} di Roma,
Piazzale Aldo Moro 5, I-00185 Roma, Italy}

\author{Chiara Vitelli}
\affiliation{Dipartimento di Fisica, Sapienza Universit\`{a} di Roma,
Piazzale Aldo Moro 5, I-00185 Roma, Italy}
\affiliation{Center of Life NanoScience @ La Sapienza, Istituto
Italiano di Tecnologia, Viale Regina Elena, 255, I-00185 Roma, Italy}

\author{Fulvio Flamini}
\affiliation{Dipartimento di Fisica, Sapienza Universit\`{a} di Roma,
Piazzale Aldo Moro 5, I-00185 Roma, Italy}

\author{Niko Viggianiello}
\affiliation{Dipartimento di Fisica, Sapienza Universit\`{a} di Roma,
Piazzale Aldo Moro 5, I-00185 Roma, Italy}

\author{Ludovico Latmiral}
\affiliation{Dipartimento di Fisica, Sapienza Universit\`{a} di Roma,
Piazzale Aldo Moro 5, I-00185 Roma, Italy}

\author{Paolo Mataloni}
\affiliation{Dipartimento di Fisica, Sapienza Universit\`{a} di Roma,
Piazzale Aldo Moro 5, I-00185 Roma, Italy}

\author{Daniel J. Brod}
\affiliation{Perimeter Institute for Theoretical Physics, 31 Caroline Street North, Waterloo, ON N2L 2Y5, Canada}

\author{Ernesto F. Galv\~{a}o}
\affiliation{Instituto de F\'isica, Universidade Federal Fluminense, 
Av. Gal. Milton Tavares de Souza s/n, Niter\'oi, RJ, 24210-340, Brazil}

\author{Andrea Crespi}
\affiliation{Istituto di Fotonica e Nanotecnologie, Consiglio
Nazionale delle Ricerche (IFN-CNR), Piazza Leonardo da Vinci, 32,
I-20133 Milano, Italy}
\affiliation{Dipartimento di Fisica, Politecnico di Milano, Piazza
Leonardo da Vinci, 32, I-20133 Milano, Italy}

\author{Roberta Ramponi}
\affiliation{Istituto di Fotonica e Nanotecnologie, Consiglio
Nazionale delle Ricerche (IFN-CNR), Piazza Leonardo da Vinci, 32,
I-20133 Milano, Italy}
\affiliation{Dipartimento di Fisica, Politecnico di Milano, Piazza
Leonardo da Vinci, 32, I-20133 Milano, Italy}

\author{Roberto Osellame}
\affiliation{Istituto di Fotonica e Nanotecnologie, Consiglio
Nazionale delle Ricerche (IFN-CNR), Piazza Leonardo da Vinci, 32,
I-20133 Milano, Italy}
\affiliation{Dipartimento di Fisica, Politecnico di Milano, Piazza
Leonardo da Vinci, 32, I-20133 Milano, Italy}

\author{Fabio Sciarrino}
\email{fabio.sciarrino@uniroma1.it}
\affiliation{Dipartimento di Fisica, Sapienza Universit\`{a} di Roma,
Piazzale Aldo Moro 5, I-00185 Roma, Italy}

\maketitle

\textbf{Boson Sampling is a computational task strongly believed to be hard for classical computers, but efficiently solvable by orchestrated bosonic interference in a specialised quantum computer.  Current experimental schemes, however, are still insufficient for a convincing demonstration of the advantage of quantum over classical computation. A new variation of this task, Scattershot Boson Sampling, leads to an exponential increase in speed of the quantum device, using a larger number of photon sources based on parametric downconversion. This is achieved by having multiple heralded single photons being sent, shot by shot, into different random input ports of the interferometer. Here we report the first Scattershot Boson Sampling experiments, where six different photon-pair sources are coupled to integrated photonic circuits. We employ recently proposed statistical tools to analyse our experimental data, providing strong evidence that our photonic quantum simulator works as expected. This approach represents an important leap toward a convincing experimental demonstration of the quantum computational supremacy.}\\
\section*{Introduction}
\begin{figure*}
\centering
\includegraphics[width=0.94\textwidth]{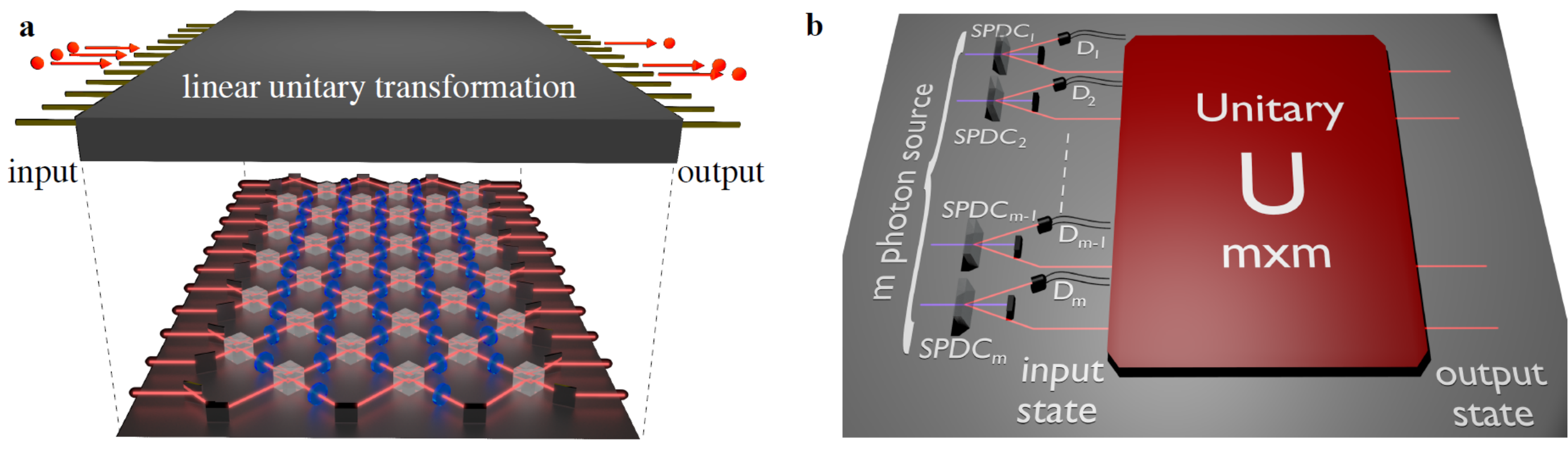}
\caption{{\bf Boson Sampling and its Scattershot configuration}. {\bf a}, Conceptual scheme of Boson Sampling with $n$ bosons undergoing an arbitrary $m$-mode unitary transformation. The problem is to sample from the output distribution of the $n$-bosons over the $m$-modes. This task can be efficiently performed by a specialized quantum computer performing $n$-photon interference in a $m$-mode linear interferometer implementing the chosen unitary transformation. {\bf b}, Scattershot configuration for Boson Sampling with randomly chosen inputs. $m$ heralded single-photon sources, one for each input port, are coupled to the interferometer. During a given time period, $n$ photons ($n<m$) are probabilistically injected into the interferometer. Each detected $n$-photon event at the interferometer's output can be assigned to its corresponding input state by the heralding detectors. Boson Sampling is thus performed with random, but heralded, inputs \cite{Lund2013,AaronsonBlog}.
}
\label{fig:concept}
\end{figure*}

Theory has shown that quantum computers should be able to markedly outperform conventional, classical computers in specific tasks \cite{Preskill12}. In practice, however, no quantum computer has yet solved a problem instance which is hard to solve classically. With the goal of rigorously establishing what was called quantum supremacy, in 2010 Aaronson and Arkhipov provided strong theoretical evidence that a simpler, specialised quantum computer could solve a classically-hard computational task \cite{Aaronson10}. The so-called Boson Sampling problem consists in sampling from the output distribution of $n$ indistinguishable photons entering different input modes of a given $m$-mode random interferometer (see Fig. \ref{fig:concept}a). The complex multi-photon interference within the device was shown, under mild computational assumptions, to yield an output distribution that is hard to sample using classical computers. The difficulty has been traced back to the known intractability of calculating the permanent function of a matrix \cite{Valiant79}. Indeed, each output's probability amplitude is given by the permanent of a different $n \times n$ matrix obtained from the $m \times m$ unitary matrix $U$ describing the interferometer \cite{Aaronson10,Troyansky96,Scheel04}.\\
Because a photonic Boson Sampling computer does not use adaptive measurements, it falls short of the requirements \cite{Ladd2010,KLM01} for a universal quantum computer capable, for example, of factoring integers efficiently \cite{Shor97}. On the other hand, its comparatively simple design has prompted a number of small-scale implementations using the interference of 3 photons injected over different modes in integrated interferometers with up to 13 modes \cite{Broome2013,Spring2013,Till2012,Crespi2012,Spagnolo2013,Spagnolo2013a,Carolan2013a}. First estimates have shown that 30 photons evolving in an interferometer with about 100 modes would already be extremely demanding to simulate classically, providing strong experimental evidence for the quantum computational supremacy. Moreover, Boson Sampling is an experimental platform suitable for addressing important intermediate challenges for the field of quantum computation, such as benchmarking and certification of medium-scale devices \cite{Gogolin2013, Aaronson13, Spagnolo2013a,Carolan2013a}. There have been recent theoretical investigations on allowable error tolerances \cite{Rohde12,Leverrier2013} as well as a recent proposal for an implementation using phonons in ion traps \cite{Lau2012,Shen2013}. The technologies enabling a Boson Sampling computer are useful also for other photonic applications such as quantum cryptography \cite{RevModPhys.74.145} and universal photonic quantum computation \cite{KLM01,Kok2007}.\\
One of the main difficulties in scaling up the complexity of Boson Sampling devices is the requirement of a reliable source of many indistinguishable photons. Despite recent advances in photon generation \cite{Eisaman2011} using atoms \cite{atoms2}, molecules \cite{molecules3,molecules2}, colour centers in diamond \cite{color} and quantum dots \cite{Qdot3,Qdot4}, currently the most widely used method remains parametric downconversion (PDC) \cite{SPDC1,SPDC2}. This approach requires pumping a nonlinear crystal with an intense laser to generate pairs of identical photons. The main advantages of PDC sources are the high photon indistinguishability, collection efficiency and relatively simple experimental setups. This technique, however, suffers from two drawbacks. First, because the nonlinear process is non-deterministic, so is the photon generation, even though it can be heralded. Second, the laser pump power, and hence the source's brilliance, has to be kept low to prevent unwanted higher-order terms in the photon generation process. These two characteristics have, so far, restricted PDC implementations of Boson Sampling experiments to proof-of-principle demonstrations with 3 photons only in the original spirit of Boson Sampling (one photon per mode, injected over different modes).\\
Recently, a new scheme has been proposed to make the best use of PDC sources for photonic Boson Sampling, greatly enhancing the rate of $n$-photon events \cite{Lund2013,AaronsonBlog}. This approach has been named Scattershot Boson Sampling in Aaronson's blog\cite{AaronsonBlog} and involves connecting $k$ ($k>n$) PDC heralded single-photon sources to different input ports of the interferometer (see Fig. \ref{fig:concept}b). Suppose each PDC source yields a single photon with probability $\epsilon$ per pulse. By pumping all $k$ PDC crystals with simultaneous laser pulses, $n$ photons will be simultaneously generated in a random (but heralded) set of input ports with probability $\binom{k}{n} \epsilon^{n}$, which for $k \gg n$ represents an exponential improvement in generation rate with respect to usual, fixed-input Boson Sampling with $n$ sources. The Scattershot Boson Sampling problem, naturally solvable by this setup, is to sample from the output distribution of a given, random interferometer for random sets of input modes.
Note that the pump laser power does not need to be increased $k$-fold, as the laser can sequentially pump each PDC source with very little loss to down-converted photons. In this way, the ratio between one-pair production rate and higher order terms can be kept low. Another interesting feature of this scheme is the possibility of recording events corresponding to different numbers of injected photons. All these characteristics suggest that the Scattershot approach to Boson Sampling will be decisive in future, larger experiments designed to reach the quantum supremacy regime.\\
\begin{figure*}
\centering
\includegraphics[width=0.99\textwidth]{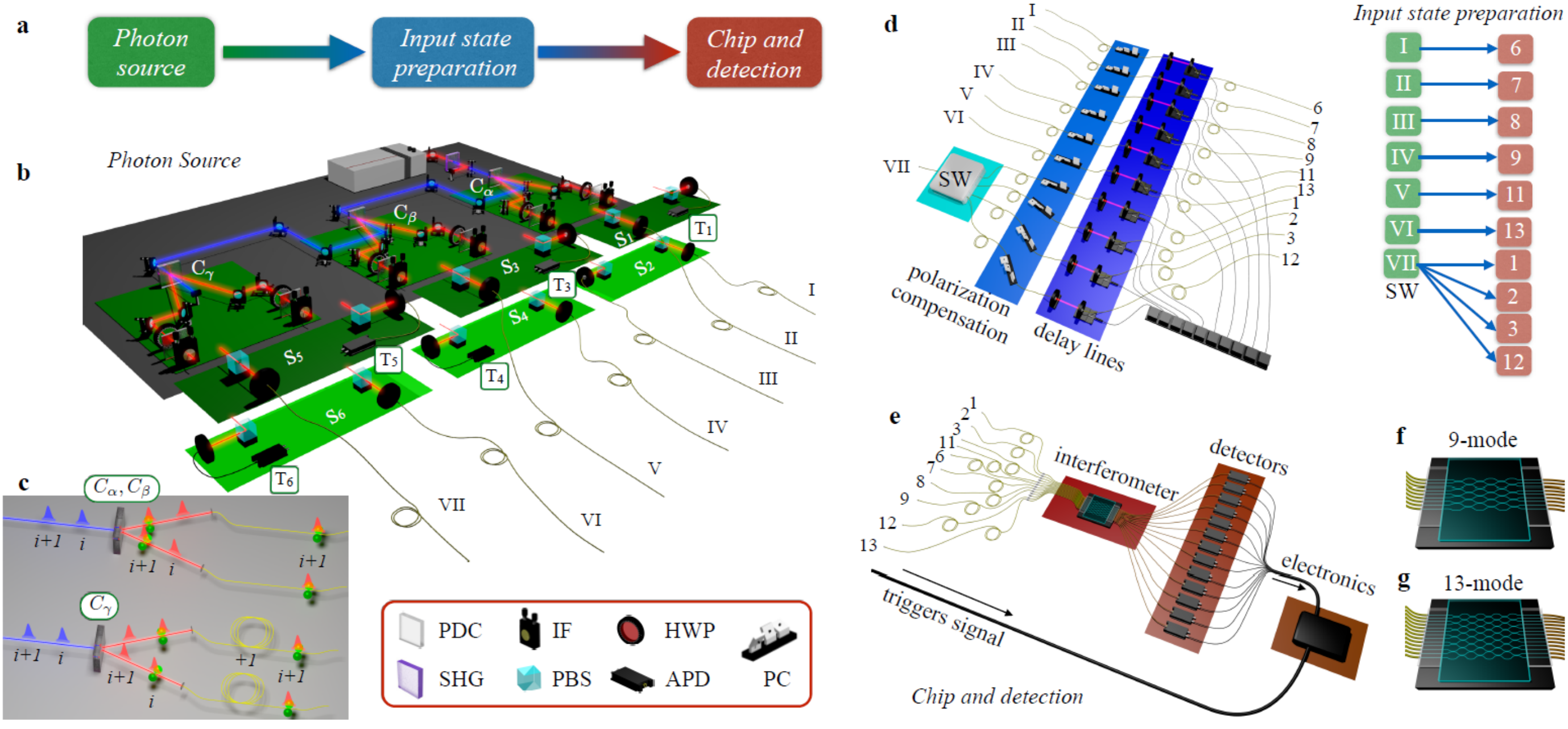}
\caption{{\bf Experimental layout for the implementation of Boson Sampling with multiple inputs.} {\bf a}, Overall conceptual scheme of the experiment. {\bf b}, In each of the 3 BBO crystals ($C_\alpha$, $C_\beta$ and $C_\gamma$), photon pairs are generated via type II parametric down conversion (PDC) process. The two possible polarization combinations for the two generated photons, HV and VH, constitute two equal PDC sources enfolded in the same crystal, each one exciting a different trigger (photon V) and a different input mode (photon H). The only exception is given by source $S_2$, whose outputs (I and III in the figure) are both injected in the chip. Sources are also time-multiplexed, since pulses generating photons in crystal $C_\gamma$ are produced before the ones generating photons in $C_\alpha$ and $C_\beta$. {\bf c}, Schematic visualization of the six time- and space-multiplexed photon sources; $i$ is an index of the pump pulse number. {\bf d}, For the 9-mode device, the input state is varied manually by changing the input fibers. For the 13-mode device, the input state is varied by the multiple source configuration and by the photon switcher, as described in the main text. Top right inset: map of the connections between sources and interferometer's inputs. {\bf e}, The photons are then injected into the interferometer by means of a single-mode fiber-array and then collected at the output via a multi-mode fiber-array, connected to a set of avalanche photodiodes for detection.  {\bf f-g}, Internal waveguide design of the 9-mode ({\bf f}) and 13-mode interferometers ({\bf g}). Directional couplers have transmittivity $t^{2}_i=0.5$, while the interferometer's structure presents static phase shifts with a random pattern. Legend: SHG - Second Harmonic Generation; PDC - Parametric Down Conversion; HWP - Half Wave Plate; IF - Interference Filter; PBS - Polarizing Beam Splitter; APD - Avalanche Photodiode; PC - Polarization Controller; SW - Fiber Switcher.}
\label{fig:expsetup}
\end{figure*}
\begin{figure*}
\centering
\includegraphics[width=0.99\textwidth]{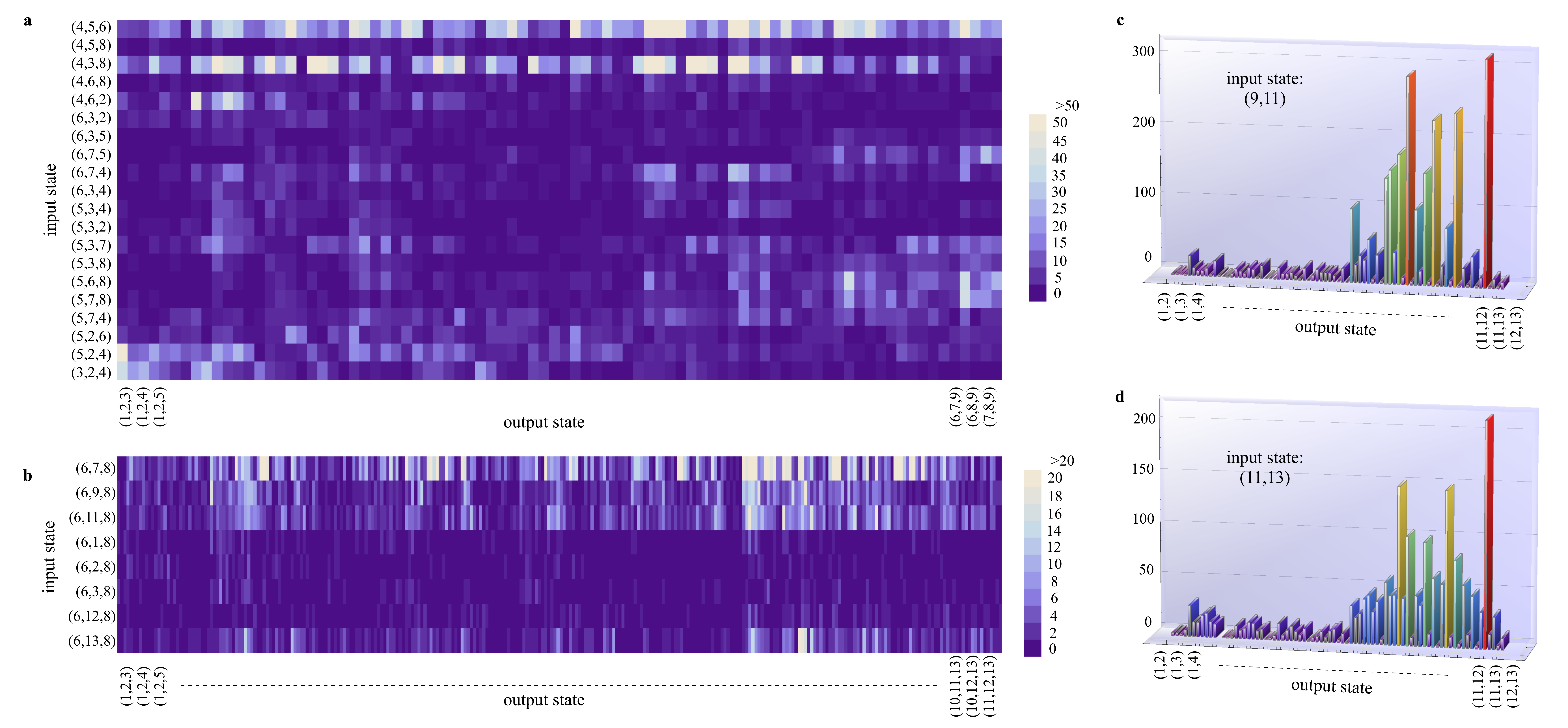}
\caption{{\bf Multiple input Boson Sampling in a 9-mode device and Scattershot Boson Sampling in a 13-mode device.} {\bf a}, Density plot of the number $n_{i,j}$ of events detected for each of the 1680 input ($i$)  and output ($j$) combinations used in our  Boson Sampling experiments with the $9$-mode chip. {\bf b}, Density plot of the number $n_{i,j}$ of events detected for each of the 2288 input ($i$)  and output ($j$) combinations used in our Scattershot Boson Sampling experiment with the $13$-mode chip. {\bf c,d}, Number $n_{i,j}$ of events detected for a two-photon Scattershot experiment with the $13$-mode chip for input states (9,11) ({\bf c}) and (11,13) ({\bf d}).}
\label{fig:dataanalysis1}
\end{figure*}
Here we report experimental results on Scattershot Boson Sampling experiments using a $13$-mode integrated photonic chip. We use up to six PDC photon sources to obtain data corresponding to 2- and 3-photon interference, and validate the device's functioning using recently proposed statistical tests \cite{Aaronson13, Spagnolo2013a}. Additional results on a different 9-mode chip are also presented and certified, thus showing the robustness of the Scattershot approach. Finally, we use numerical calculations to discuss the complexity of Boson Sampling simulation and certification, and to estimate a benchmark for quantum supremacy.\\
\section*{Scattershot Boson Sampling experiment}
A photonic Scattershot Boson Sampling experiment involves a few experimentally demanding steps (see Fig. \ref{fig:expsetup}a). First, $k>n$ PDC sources are used to generate $n$ indistinguishable photons in a heralded, but random, set of modes. The input state must then be prepared (introducing time delays and polarisation compensation) to be injected into the $m$-mode integrated interferometer. We must then detect $n$-fold coincident photon counts at the chip's output modes, all the while maintaining synchronisation so that we have true $n$-photon interference in the chip. Finally, it is necessary to analyse the output data to validate the correct functioning of the device.\\
For our experiments we fabricated two integrated photonic chips implementing random multimode interferometers (with 9 and 13 modes), using a femtosecond laser writing technique \cite{gattass2008flm,Marshall09,corrielli2014rotated,heilmann2014arbitrary} described in the Methods section. For the 9-mode chip, the input state was created by a 4-photon PDC source (crystal $C_{\alpha}$ in Fig. \ref{fig:expsetup}b), with one of the photons used as a trigger. Our preliminary experiment involved simulating the statistics of a Scattershot Boson Sampling experiment in the 9-mode chip by manually connecting 20 different sets of input modes to the source, via a fiber array, and uniformly mixing the data corresponding to different input states.\\
We used the 13-mode chip to implement Scattershot Boson Sampling experiments with a total of six PDC sources ($S_1$ to $S_6$ in Fig. \ref{fig:expsetup}b). We simplified the implementation by enfolding two equal sources in each crystal, corresponding to the two possible vertical/horizontal polarization combinations for the photon pair generated. Hence, the six sources $S_1$-$S_6$ are created using only the three crystals $C_\alpha$, $C_\beta$ and $C_\gamma$.  Each PDC source ideally produces two indistinguishable photons. One such source (Source $S_2$) prepares photons I and III, which enter the interferometer in fixed modes 6 and 8, respectively. The other five PDC sources produce random, but heralded, single photons, which are coupled to different input ports of the chip via a polarisation correction stage, delay lines, and a single-mode fiber array, according to the map in Fig. \ref{fig:expsetup}d. Note that we further increased the input variability by distributing photon VII randomly among four different input ports, via an optical fiber switcher with switching rate comparable to the obtained experimental count rate. This raises from five to eight the number of possible input sets, allowing us validation procedure tests on data sampled from a larger number of input-output configurations.\\
For both chips, the output photons are collected by a multimode fiber array, and multiphoton coincidences are detected by avalanche photodiodes, coordinated by an electronic data acquisition system capable of registering events with an arbitrary number of photons. We then analyzed data corresponding to 2- and 3-photon interference inside the chip. Synchronizing up to six PDC sources distributed over 10 input modes is a technically difficult step; once that was achieved, the controllable, relative delays between photons allowed us to adjust their degree of distinguishability. Further details about synchronization procedures and indistinguishability between photons of different sources are given in the Supplementary Materials.\\
The observed number of events corresponding to each input/output combination for the 9- and 13-mode chips are shown respectively in Figs. \ref{fig:dataanalysis1}a and \ref{fig:dataanalysis1}b. 
Note the sparseness of the data set, as only a few events corresponding to each input/output combination are observed (if any). This is an expected feature of more complex Boson Sampling experiments whose number of possible input-output combinations may far exceed the number of observed events. Furthermore, in Figs. \ref{fig:dataanalysis1}c and \ref{fig:dataanalysis1}d we show the results for 2-photon experiments, in which each input is a doubly-heralded 2-photon state.\\
Another route to more complex Boson Sampling experiments is time-multiplexing \cite{Pittman,Jeffrey,Migdall,McCusker,Zeilinger}, that is, exploiting interference of photons created by different pump pulses on the same PDC source. Ultra-fast optical switchers can be used to distribute the photons generated by subsequent pump pulses to different input ports of the photonic chip, after suitable synchronisation delays. This type of time-multiplexing increases the $n$-photon generation, using a fixed number of PDC sources. Our experiments with the 13-mode chip feature a first proof-of-principle demonstration of interference among photons generated by different pulses. This was done by introducing appropriate delays, so that photons from Sources $S_5$ and $S_6$ are produced by a different pump pulse than those generated by all the other sources (see Fig. \ref{fig:expsetup}c).

\begin{figure*}
\centering
\includegraphics[width=0.95\textwidth]{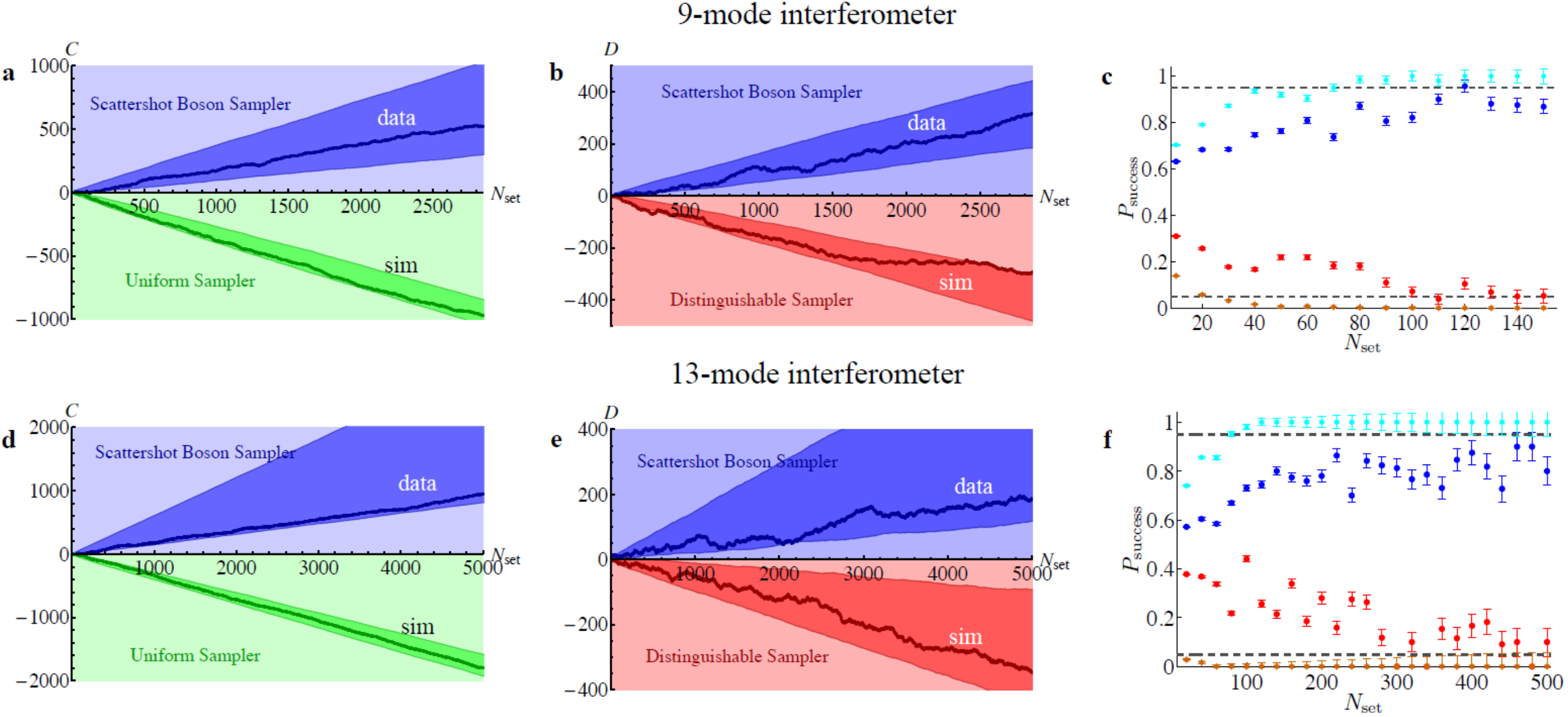}
\caption{{\bf Validation of multiple-input and Scattershot Boson Sampling against various alternative distributions.}
{\bf a,d} Application of the Aaronson and Arkhipov test against the Uniform distribution ({\bf a}, for the 9-mode chip and {\bf d}, for the 13-mode chip). {\bf b,e} Application of the likelihood ratio test against Distinguishable Sampler ({\bf b}, for the 9-mode chip and {\bf e}, for the 13-mode chip). {\bf c,f} Success probability $P_{\mathrm{success}}$ of the validation protocol against different alternative distributions as a function of the data set size $N_{\mathrm{set}}$ ({\bf c}, for the 9-mode chip and {\bf f}, for the 13-mode chip). Horizontal dashed line: $0.95$ and $0.05$ thresholds for the success probability $P_{\mathrm{success}}$.
{\it Legend for panels} {\bf a-b, d-e}. Blue points: Scattershot Boson Sampling experimental data. Green points: numerical simulation of a Uniform Sampler. Red points: numerical simulation of Distinguishable Sampler data. Dark blue areas: $\pm 2 \sigma$ region ({\bf a,d}) or $\pm 1 \sigma$ region ({\bf b,e}) expected for the experimental Scattershot data, obtained from a numerical simulation which includes noise in the implemented unitary corresponding to the fabrication tolerances. Dark green areas: $\pm 2 \sigma$ region expected for the Uniform Sampler. Dark red areas: $\pm 1 \sigma$ region expected for the Distinguishable Sampler.
{\it Legend for panels} {\bf c,f}. Cyan points: Scattershot Boson Sampling experimental data against the Uniform Sampler with the Aaronson-Arkhipov test. Blue points: Scattershot Boson Sampling experimental data against the Distinguishable Sampler. Orange points: numerical simulation of Uniform Sampler data against Scattershot Boson Sampler. Red points: numerical simulation of Distinguishable Sampler data against the Scattershot Boson Sampler.}
\label{fig:dataanalysis2}
\end{figure*}

\subsection*{Validation of experimental Boson Sampling data}

Unlike problems such as integer factoring, the full certification of the correct functioning of a Boson Sampling device is by itself a hard computational problem \cite{Aaronson10,Gogolin2013,Aaronson13,Aolita14,PhysRevLett.113.020502}. There are, however, statistical tests able to provide partial certification against a number of sensible hypotheses about how the device may be failing to sample from the correct, ideal distribution. Boson Sampling thus serves as a useful test bench for the more general problem of quantum device certification. We now discuss the results of the application of validation tests designed for standard Boson Sampling experiments to our Scattershot scenario.\\
The first test we applied to our data is the scalable statistical test proposed by Aaronson and Arkhipov in Ref. \cite{Aaronson13}, initially designed to distinguish fixed-input Boson Sampling events from a uniform distribution over the possible outputs and here extended to the Scattershot scenario. This is achieved by calculating, for each observed event, a discriminator $P$ which weakly correlates with the Boson Sampling probability, but which can be calculated efficiently \cite{Spagnolo2013a}. The result for the 9-mode chip is reported in Fig. \ref{fig:dataanalysis2}a; at variance with the test performed in Ref. \cite{Spagnolo2013a}, instead of a single input our 9-mode chip experiments allowed for 1680 different input-output combinations. We have also applied the test to data obtained from the 13-mode chip, and the results are reported in Fig. \ref{fig:dataanalysis2}d; in this case, there were 2288 different input/output combinations.\\
A second test we performed is an adaptation of a standard likelihood ratio test \cite{CoverThomas06}, with the goal of comparing our experimental data with those expected if distinguishable photons were used. For each experimental outcome, the probabilities for indistinguishable and distinguishable photons are compared (more details on the tests are reported in the Methods section and in the Supplementary Materials). The results of this test for the 9-mode and 13-mode chips are shown respectively in Figs. \ref{fig:dataanalysis2}b and \ref{fig:dataanalysis2}e. Note that, again, in both cases we applied the test to the data set combining all different input states used.\\
Successful validation could be obtained even with small data sets. This is highlighted in Figs. \ref{fig:dataanalysis2}c for the 9-mode chip and \ref{fig:dataanalysis2}f for the 13-mode chip, where we plot the trend of the test's success rate against the size of the data set used.

\begin{figure*}
\centering
\includegraphics[width=0.95\textwidth]{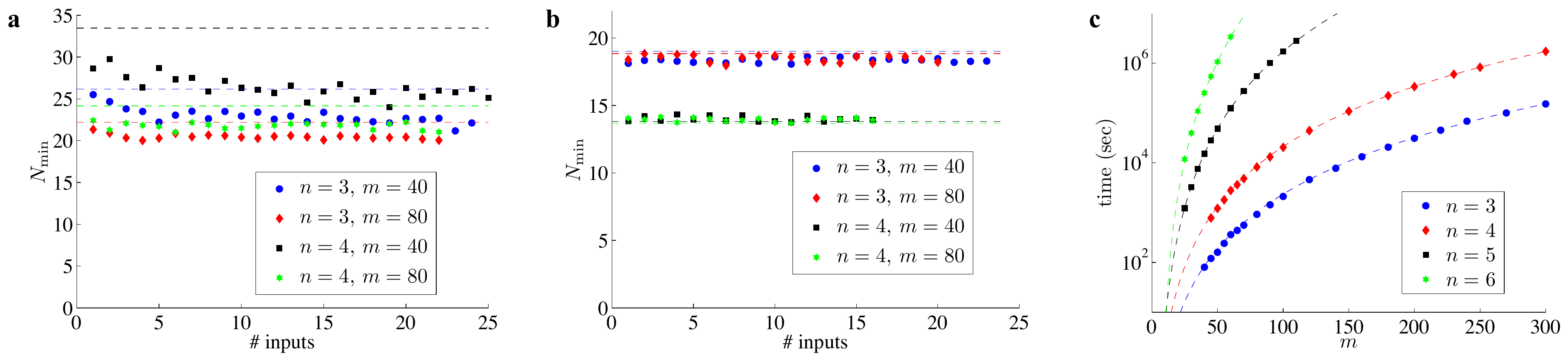}
\caption{{\bf Full simulation of Scattershot Boson Sampling and of its validation}. {\bf a} Minimum data set size to obtain $95\%$ success probability for the validation of Scattershot Boson Sampling data against the Uniform Sampler as a function of the number of input states, adopting the Aaronson and Arkhipov test. {\bf b}, Minimum data set size to obtain $95\%$ success probability for the validation of a Scattershot Boson Sampling experiment against the Distinguishable Sampler as a function of the number of input states, adopting the likelihood ratio test. {\bf c}, Time required with a laptop to calculate $N=2000$ Boson Sampling probability distributions, each one corresponding to a different input configuration, as a function of the number of modes $m$, for different number of photons $n$. 
}
\label{fig:compute}
\end{figure*}

\section*{Discussion}

In summary, we have reported the first experimental implementation of the Scattershot approach to photonic Boson Sampling, recently proposed in \cite{Lund2013,AaronsonBlog}, a promising way of exponentially scaling up the computational power of the quantum sampler. 
Our experiments use 6 PDC sources in parallel to demonstrate the feasibility of non-trivial realizations of this approach. In the experimental implementation, due to non-optimal beam propagation and PDC sources, we observed an increase in the event rate by a factor 4.5 (3.4), compared to standard Boson Sampling with a source of average (best) brightness. This value should be compared to the expected value of 5.\\
Let us now discuss how Scattershot Boson Sampling may bring within reach an experimental regime  approaching quantum supremacy. Let us consider experiments with $N = 2000$ events, more than sufficient to perform a successful validation of the Scattershot Boson Sampler (see Figs. \ref{fig:compute}a and \ref{fig:compute}b). In this regime, with high probability each recorded event is sampled from a different input state, provided that $\binom{m}{n} \gg N$, and assuming the use of one PDC source per input mode. To get an insight into the hardness of calculating the whole output distribution corresponding to each input used, we illustrate the required computational time on a standard laptop in Fig. \ref{fig:compute}c. Although this brute force calculation is currently the only reported approach for a classical Boson Sampling simulation, it is likely that more efficient classical sampling algorithms are possible for interferometers chosen uniformly at random, but no description of those has yet been reported in the literature \cite{PrivateComm}.\\
The main advantage of the Scattershot approach is to markedly decrease the experimental run-time with respect to the usual, fixed-input Boson Sampling setup. Using challenging but feasible experimental parameters for pulse rate (80 MHz), per-pulse generation probability (0.015), triggering efficiency (0.5) and overall photon counting probability (0.15, which takes into account both photon losses in the injection-propagation stage, linearly dependent from the chip size, and detector inefficiencies), we get an estimated runtime of $\sim 10^7-10^8$ seconds for a $2000$-event, fixed-input Boson Sampling experiment with $n=4$, $m=100$. The corresponding Scattershot Boson Sampling experiment uses $k=100$ PDC sources in parallel, resulting in a quantum runtime of $\sim 50$ seconds. These estimates clearly illustrate the boost in computational speed provided by the Scattershot approach.\\
Validation of the Scattershot Boson Sampler would still be feasible well into the quantum supremacy regime, as the number of events whose probabilities need to be calculated by a classical computer to certify the proper operation of the quantum device is very low and almost independent on the number of photons and modes involved (see Figs. \ref{fig:compute}a and \ref{fig:compute}b). This is expected to hold for validation of experiments with up to about 30 photons.\\
Note that the simulations of Fig. \ref{fig:compute} did not take into account errors such as partial photon distinguishability and other experimental imperfections. In larger devices, for example, a spurious but genuine-looking event could result from the loss of $l<n$ triggered photons and simultaneous injection of $l$ untriggered photons. A precise analysis of the effect of incorrectly heralded photons in our experiments is carried out in the Supplementary Materials. These events count as white noise in the validation tests, slightly lowering the test's efficiency (see the Supplementary Materials section for more details). This particular problem can be overcome by using the heralding detectors to briefly open an optical shutter in the corresponding input mode, as discussed in the Supplementary Materials.\\
Other photon source schemes, such as collecting larger number of modes from degenerate PDC type I radiation via microlenses \cite{Rossi09}, as well as novel approaches using time-bin encoding \cite{Motes14} are all promising routes to scale up the complexity of future Boson Sampling experiments. Further theoretical progresses could also help in this endeavor, such as the development of scalable statistical validation tests against other alternative distributions. Recent proposals along these lines are based on looking at global coalescence effects \cite{Carolan2013a}, checking specific output suppressions in interferometers with certain symmetries \cite{PhysRevLett.113.020502}, or performing single-mode homodyne detection \cite{Aolita14}. Moreover, it has been argued that there are other classes of quantum states that can be used for Boson Sampling without spoiling its computational complexity \cite{Dowling1,Sesha2014}; future research in this direction could help to simplify the experimental implementation of hard-to-simulate devices.

\section*{Methods}

\subsection*{Fabrication of integrated optics devices} Multimode integrated interferometers are fabricated in Eagle2000 (Corning) alumino-borosilicate glass by femtosecond laser direct writing. Focused ultrashort pulses induce permanent refractive index changes in the focal volume by nonlinear absorption mechanisms. Buried waveguides are directly drawn in the volume of the glass by suitably translating the sample with respect to the writing beam. This direct-write technique allows fast realization of custom integrated optical circuits with large design freedom.  A cavity dumped Yb:KYW mode-locked oscillator, producing laser pulses with $\sim$300~fs duration, 1~MHz repetition rate and 1030~nm wavelength, is employed. In particular, irradiation is performed by focusing 220 nJ pulses with a 0.6~NA microscope objective and by translating the sample at 40~mm/s constant speed, to obtain single-mode waveguides for 785~nm photons. Average waveguide depth below the sample surface is 170~$\mu$m. Interferometers implementing random unitary matrices are obtained by cascading several rows of balanced (50:50) directional couplers, with the layouts in Figs. \ref{fig:expsetup}f and \ref{fig:expsetup}g, connected by S-bends of slightly different lengths, which induce controlled (though randomly chosen) phase shifts \cite{Crespi2012}. Each directional coupler (including S-bends) is about 5~mm long, while input and output waveguides are 127~$\mu$m spaced, for a global footprint of the circuits of about 35~mm~$\times$~1.1~mm for the 9-mode device and 45~mm~$\times$~1.6~mm for the 13-mode device.\\
\subsection*{Experimental details}
Single photons were generated in six equal parametric down conversion (PDC) sources, implemented in three crystals. The 3-photon input state for the 9-mode chip was obtained by PDC generation from the first crystal, with one of the four emitted photons used as a trigger. The input states were then changed manually by connecting a fiber array to 20 different sets of input modes of the chip. The 13-mode chip was then used to implement the complete scattershot version of the Boson Sampling experiment. The three crystals reproduced six PDC sources, the first one belonging to the first crystal was adopted to inject two fixed input modes of the chip (number 6 and number 8), while another photon was injected shot by shot coming from one of the five remaining PDC sources. At the output of both chips, multimode fibers were connected to single photon counting detectors and an electronic data acquisition system allowed to register events with an arbitrary number of photons.\\
\subsection*{Validation of the experimental data} The validation against the hypothesis that the data are sampled according to a uniform distribution is performed by adopting the scalable Aaronson and Arkhipov test \cite{Aaronson13} experimentally verified in \cite{Spagnolo2013a}. The validation test against the hypothesis that the data are sampled with distinguishable photons works as follows. For each experimental outcome $i$, the certifier calculates the associated probabilities $p_{i}^{\mathrm{ind}}$ for indistinguishable photons and $q_{i}^{\mathrm{dis}}$ for distinguishable photons. A counter variable $D$ is increased (decreased) by 1 if $p_{i}^{\mathrm{ind}} > q_{i}^{\mathrm{dis}}$ ($p_{i}^{\mathrm{ind}} < q_{i}^{\mathrm{dis}}$). After analysing all events, $D>0$ ($D<0$) indicates the hypothesis of indistinguishable (distinguishable) photons is more likely to hold. The probabilities $p_i$ and $q_i$ are calculated using the permanent formula, taking into account the partial photon distinguishability of the source, and the chip's theoretical design parameters. For the 9-mode interferometer, the data were collected separately by manually changing the input state. Then, the recorded events before the validation procedure are mixed uniformly in order to represent a set of data collected with a random input state.\\
The same validation procedure was carried out for the 2-photon data, which were collected simultaneously to the 3-photon ones. In particular, photons from inputs 11 and 13 are generated from two different laser pulses. We obtained an average success probability $P_{\mathrm{success}} > 95\%$ of the validation process after a data set size of $N_{\mathrm{set}} \sim 150$ against the Uniform distribution and of $N_{\mathrm{set}} \sim 50$ against the distribution with distinguishable photons.


This work was originally published as Bentivegna et al. Sci. Adv. 2015; 1:e1400255, http://advances.sciencemag.org/content/1/3/e1400255. The full legal code of the CC BY-NC Public License may be found at http://creativecommons.org/licenses/by-nc/4.0/legalcode.

{\bf Acknowledgements}. We acknowledge extremely useful and stimulating discussion with Scott Aaronson. We acknowledge technical support from Sandro Giacomini and Giorgio Milani. This work was supported by the ERC-Starting Grant 3D-QUEST (3D-Quantum Integrated Optical Simulation; grant agreement no. 307783): http://www.3dquest.eu, by the PRIN project Advanced Quantum Simulation and Metrology (AQUASIM), by the H2020-FETPROACT-2014 Grant QUCHIP (Quantum Simulation on a Photonic Chip; grant agreement no. 641039) and by the Brazilian National Institute for Science and Technology of Quantum Information (INCT-IQ/CNPq). D.J.B. was supported in part by Perimeter Institute for Theoretical Physics. 

M.B., N.S., C.V., F.F., P.M., E.F.G., R.O., and F.S. conceived the experimental implementation of the Scattershot Boson Sampling. A.C., R.R. and R.O. fabricated and characterized the integrated devices using classical optics. C.V., M.B., N.S., F.F., N.V. and F.S. carried out the quantum experiments. N.S., M.B., C.V., F.F. and F.S. elaborated the data. N.S., M.B., L.L., C.V., F.F., D.B., E.F.G., and F.S. carried out the numerical simulation. All the authors discussed the experimental implementation and results, and contributed to writing the paper.

\end{document}